\documentclass[pre,epsfig,showpacs,address,onecolumn]{revtex4}%
\usepackage[english]{babel}
\usepackage[latin1]{inputenc}
\usepackage[dvips]{graphicx}
\usepackage{amsmath}
\usepackage{amssymb}
\usepackage{epsfig}
\usepackage{array}
\usepackage{amsfonts}
\usepackage{graphicx}
\usepackage{subfigure}%
\begin{document}
\title{Driven isotropic Heisenberg spin chain with arbitrary boundary twisting angle:
exact results}
\author{V.~Popkov$^{~1,2}$, D. Karevski$^{3}$, G.M.~Sch\"utz$^{4}$}
\affiliation{$^{1}$ Dipartimento di Fisica e Astronomia, Universit\`a di Firenze, via G.
Sansone 1, 50019 Sesto Fiorentino, Italy}
\affiliation{$^{2}$ Institut f\"ur Theoretische Physik, Universit\"at zu K\"oln,
Z\"ulpicher Str. 77, D-50937 Cologne, Germany}
\affiliation{$^{3}$ Institut Jean Lamour, dpt. P2M, Groupe de Physique Statistique,
Universit\'e de Lorraine, CNRS, B.P. 70239, F-54506 Vandoeuvre les Nancy
Cedex, France}
\affiliation{$^{4}$Institute of Complex Systems II, Forschungszentrum J\"ulich, 52428
J\"ulich, Germany}
\date{\today }

\begin{abstract}
We consider an open isotropic Heisenberg quantum spin chain, coupled at the
ends to boundary reservoirs polarized in different directions, which sets up a
twisting gradient across the chain. Using a matrix product ansatz, we
calculate the exact magnetization profiles and magnetization currents in the
nonequilibrium steady steady state of a chain with $N$ sites. The
magnetization profiles are harmonic functions with a frequency proportional to
the twisting angle $\theta$. The currents of the magnetization components
lying in the twisting plane and in the orthogonal direction behave
qualitatively differently: In-plane steady state currents scale as $1/N^{2}$
for fixed and sufficiently large boundary coupling, and vanish as the coupling
increases, while the transversal current increases with the coupling and
saturates to $2\theta/N$.

\end{abstract}

\maketitle

%pacs: {Open systems, Spin chain models,Integrable systems, Quantum transport}

%\pacs{87.14.Ee, 87.15.Aa, 87.15.Vv}
The Heisenberg quantum spin chain is a fundamental and one of the most
well--studied quantum models of statistical mechanics. However, not much is
known about the Heisenberg chain in a nonequilibrium setting where the chain
is maintained in a strongly non-equilibrium state by e.g. a coupling to the
ends to boundary reservoirs at different chemical potential or different
polarization. The boundary gradient drives a quantum system towards a
nonequilibrium steady state (NESS), typically characterized by various
nonvanishing currents of energy, magnetization etc. The open quantum system is
canonically described by the celebrated quantum master equation for the
reduced density matrix in the Lindblad form (Lindblad Master Equation, or LME)
\cite{Petruccione,PlenioJumps}.

The full LME evolution of an open system of $N$ spins is described by a
$2^{N}\times2^{N}$ reduced density matrix $\rho(t)$, which has $2^{2N}-1$
independent real entries. It is clear that the usage of even most powerful
numerical methods is restricted due to exponentially growing complexity of the
problem. However, many fundamental properties like conductivity, universality
classes and critical behaviour are the properties of thermodynamically large
systems and hence require the development of analytic non-perturbative
methods. In spin chain materials like SrCuO$_{2}$ many transport
characteristics are measurable experimentally
\cite{spinchain,HessBallistic2010}.

The purpose of the present paper is to investigate in detail the NESS of a
simply formulated, and analytically treatable, open quantum many-body system.
Namely, we consider an open nonequilibrium isotropic Heisenberg spin chain,
coupled to boundary reservoirs, which tend to polarize spins at the edges
along arbitrary directions $\vec{n}_{R},\vec{n}_{L}$ on the right and on the
left end, see Fig.\ref{Fig_LindbladReservoirs}. Due to the bulk isotropy, the
NESS\ depends on two scalar parameters: the angle $\theta$ between the unit
vectors $\vec{n}_{L},\vec{n}_{R},$ \ $\cos\theta=(\vec{n}_{L},\vec{n}_{R})$,
and the ratio between the boundary coupling strength and bulk exchange
interaction. Except for the case of an ideal alignment $\theta=0$, the
boundary coupling induces twisting gradient gradients and nonvanishing steady
state currents.

During last few years, the open nonequilibrium Heisenberg spin chain with
local dissipative action at the boundaries has become a paradigmatic reference
model in the field, due to its conceptual simplicity and recently discovered
powerful non-perturbative methods \cite{ProsenExact2011,MPA2013,Pros08}. The
NESS\ for the case with antiparallel alignment $\vec{n}_{L}=(0,0,1),\vec
{n}_{R}=(0,0,-1)$, so that $\theta=\pi$ was formulated as a matrix product
ansatz (MPA) and solved for a more general $XXZ$ quantum model. Further
generalizations of the basic model \cite{Pros08} were proposed, in which
additional incoherent hopping processes or bulk dephasing processes were
included \cite{EislerBallistic,Clark2013transport}.

The model with twisting boundary gradients was introduced and studied in
\cite{Lindblad2011,XYtwist} in a slightly more general setting with exchange
$Z-$anisotropy, while twisting was applied in the perpendicular $XY$-plane.
However, previous studies were limited to small system sizes, and could not
provide reliable information on scaling behaviour. Using LME symmetries
\cite{XYtwist}, one can at most make general qualitative predictions about the
nature of the steady state and admissibility of certain observables
\cite{conLivi}. E.g., based on a particular odd-even size alternating symmetry
argument, the possibility of a ballistic magnetization current in $XXZ$-model
with $XY$-twisting gradient at $\theta=\pi/2$ was ruled out \cite{XYtwist}.

In our recent study \cite{MPA2013} it was shown that the isotropic $XXX$-model
is exactly solvable by an MPA for any twisting angle $\theta$. Here we go
further and compute analytically various steady-state observables for finite
systems and in thermodynamic limit, namely the magnetization profiles and the
steady state magnetization and energy currents, as functions of the twisting
angle $\theta$ and the coupling strength $\Gamma$. Due to the isotropy
projections of the total magnetization on all three axes are individually
conserved in the steady state. We find drastically different scalings (with
system size $N$ and with coupling) of the magnetization current components
within the plane on which the twisting boundary gradient is imposed, and the
magnetization current $j_{\perp}$ in the perpendicular direction. Moreover,
for $j_{\perp}$ we find an intriguing non-commutativity of the limits
$N\rightarrow\infty$ and $\theta\rightarrow\pi$.

The plan of the paper is the following: In Sec.\ref{sec::The model} we
introduce the model and outline the Matrix Product Ansatz method. In
Sec.\ref{sec::Magnetization Profiles} we calculate the magnetization profiles,
and in Sec.\ref{sec::Magnetization Currents} the magnetization currents.
Appendix \ref{Appendix::Computation of the Z(N,theta)} contains some necessary
technical details.

\section{The model}

\label{sec::The model}

We consider an open chain of $N$ quantum spins in contact with boundary
reservoirs, the time evolution of which is given by a quantum Master equation
in the Lindblad form \cite{Petruccione,PlenioJumps}, \cite{ClarkPriorMPA2010}
(we set $\hbar=1$)%
\begin{equation}
\frac{\partial\rho}{\partial t}=-i\left[  H,\rho\right]  +4\Gamma( \mathcal{L}
_{L}[\rho]+ \mathcal{L} _{R}[\rho]). \label{LME}%
\end{equation}
Here $\rho$ is the reduced density matrix, $H$ is the isotropic spin-1/2
Heisenberg Hamiltonian
\begin{equation}
H_{XXX}=\sum_{k=1}^{N-1}\left(  \sigma_{k}^{x}\sigma_{k+1}^{x}+\sigma_{k}%
^{y}\sigma_{k+1}^{y}+\sigma_{k}^{z}\sigma_{k+1}^{z}\right)  ,
\label{Hamiltonian}%
\end{equation}
$\Gamma$ an effective coupling with the reservoirs, and $\mathcal{L} _{L}$ and
$\mathcal{L} _{R}$ are Lindblad dissipators which favour a relaxation of the
leftmost and the rightmost spins towards target states $\rho_{L},\rho_{R}$, so
that $\mathcal{L} _{L}[\rho_{L}]= \mathcal{L} _{R}[\rho_{R}]=0$. As target
states we choose fully polarized states of one spin $\rho_{L}=\frac{I}%
{2}+\frac{1}{2}\sum n_{L}^{\alpha}\sigma^{a}$, $\rho_{R}=\frac{I}{2}+\frac
{1}{2}\sum n_{R}^{\alpha}\sigma^{a}$, $\left\vert \vec{n}_{L}\right\vert
=\left\vert \vec{n}_{R}\right\vert =1$. To fix the coordinate frame, we choose
the $XY$-plane to be the plane spanned by the vectors $\vec{n}_{L},\vec{n}%
_{R}$ and the $X$-- axis to point in the $\vec{n}_{L}$ direction,
\begin{align}
\vec{n}_{L}  &  =(1,0,0)\label{nL}\\
\vec{n}_{R}  &  =(\cos\theta,\sin\theta,0), \label{nR}%
\end{align}
the angle $\theta$ between $\vec{n}_{L},\vec{n}_{R}$ taking values
$0\leq\theta\leq\pi$. A canonical, although not the most general, form of the
Lindblad action satisfying $\mathcal{L} _{L}[\rho_{L}]= \mathcal{L} _{R}%
[\rho_{R}]=0$, is
\begin{equation}
\mathcal{L} _{L,R}[\rho]=X_{L,R}\rho X_{L,R}^{+}-\frac{1}{2}\left\{
\rho,X_{L,R} ^{+}X_{L,R}\right\}  , \label{Lindblad Action}%
\end{equation}
where
\begin{equation}
\label{XLXR}X_{L}=\frac{1}{2}(\sigma^{y}+i\sigma^{z}), \quad X_{R}=\frac{1}{2}
(\sigma^{y}\cos\theta+i\sigma^{z}-\sigma^{x}\sin\theta)
\end{equation}
are polarization targeting Lindblad operators. Indeed, in absence of the
unitary term in (\ref{LME}) the boundary spins relax with a characteristic
time $\Gamma^{-1}$ to states $\rho_{L},\rho_{R}$. Consequently, the boundary
coupling introduces a twist in $XY$ plane across the whole system, which
constantly drives the system out of equilibrium. Eq. (\ref{LME}) describes the
exact time evolution of a reduced density matrix, provided that the coupling
to reservoir is rescaled appropriately with the time interval between
consecutive interactions of the system with the reservoirs, see
\cite{ClarkPriorMPA2010}.

\begin{figure}[ptb]
\begin{center}
%\centerline{\scalebox{0.6}{\includegraphics{FigLindbladReservoirsColor.eps}}}
\centerline{\scalebox{0.6}{\includegraphics{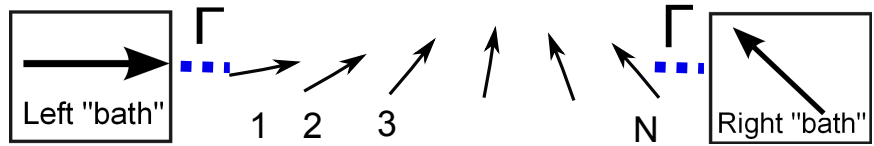}}}
\end{center}
\caption{ Schematic layout of the model. A chain of qubits is coupled at the
boundaries to the reservoirs. Dissipation introduced by the reservoirs is
described by a quantum Master equation (\ref{LME}). The figure was taken from
\cite{XYtwist}. }%
\label{Fig_LindbladReservoirs}%
\end{figure}

Our setting is shown schematically in Fig.\ref{Fig_LindbladReservoirs}. A
model with a twist (\ref{XLXR}) with right twisting angle $\theta=\pi/2$ in
$XY$ --plane and more general anisotropic $XXZ$-Hamiltonian $H_{XXZ}%
(\Delta)=\sum_{k=1}^{N-1}\left(  \sigma_{k}^{x}\sigma_{k+1}^{x}+\sigma_{k}%
^{y}\sigma_{k+1}^{y}+\Delta\sigma_{k}^{z}\sigma_{k+1}^{z}\right)  $ was
introduced in \cite{XYtwist}, and the respective NESS\ $\rho_{NESS}(\Delta)$
was shown to possess intriguing properties. The existence of a duality
transformation between $\rho_{NESS}(\Delta)$ and $\rho_{NESS}(-\Delta)$, of a
different form for even and odd $N$, allowed to conclude that the $j^{z\text{
}}$magnetization current alternates its sign with system size \cite{XYtwist}.
For large $\Gamma\gg\infty$ and for an adequate choice of the anisotropy
$\left\vert \Delta^{\ast}(N)\right\vert <1$, which depends on system size $N$,
one can generate a NESS\ arbitrarily close to a pure state, $1-Tr[\rho
_{NESS}^{2}(\Delta^{\ast})]<\varepsilon$ \cite{VNE XXZ}. For small $\Gamma
\ll1$ the NESS $\rho_{NESS}(\Delta)$ was shown to have an anomaly peaked
around the isotropic point $\Delta=1$, the anomaly turning into a singularity
at $\Delta=1$ in the limit $\Gamma\rightarrow0$ \cite{Weak XXZ}. While some of
above features were proved with general symmetry arguments, others were
conjectured by extrapolating the analytic behaviour of the system for small
sizes $N<10$ onto larger sizes. However, many important questions concerning
scaling behaviour or finite-size corrections crucial for determining the
universality, could not be answered.

In the present work we obtain exact results for the model with a boundary
twist for the thermodynamic limit $N\gg1$ by using a Matrix Product Ansatz
(MPA). Our study treats the isotropic Heisenberg chain $\Delta=1$, but an
arbitrary twisting angle $\theta$ between the targeted polarizations at the
ends. Our calculations are based on the following observation: The
unnormalized NESS of the model (\ref{LME}), (\ref{Lindblad Action}) can be
written in the form \cite{MPA2013}
\begin{align}
\rho_{NESS}  &  =\mathbf{U}S_{N}(\mathbf{U}S_{N})^{\dagger}\label{RhoNess}\\
S_{N}  &  =\left\langle 0\right\vert \Omega^{\otimes_{N}}\left\vert R_{\theta
}\right\rangle \label{S_N}\\
\mathbf{U}  &  =U^{\otimes_{N}}\nonumber
\end{align}
where $\Omega$ is $2\times2$ matrix with operator-valued entries
$\Omega(p)=S_{z}(p)\sigma^{z}+S_{+}(p)\sigma^{+}+S_{-}(p)\sigma^{-}$, which
satisfy the $SU(2)$ algebra $[S^{z},S_{\pm}]=\pm S_{\pm}$, \ $[S_{+}%
,S_{-}]=2S_{z}$ for any $p$. The representation parameter $p$ is defined by
the action of $S_{z}$ on the lowest weight vector of $SU(2)$, $\left\langle
0\right\vert S_{z}=p\left\langle 0\right\vert $, and is connected to the
coupling constant $\Gamma$ through $p=i/\Gamma$. More precisely, the operators
$S_{z}(p),S_{\pm}(p)$ operate on a semi-infinite set of states $\{\left\vert
n\right\rangle \}_{n=0}^{\infty}$, as
\begin{align*}
S_{z}(p)  &  =\sum_{n=0}^{\infty}(p-n)\left\vert n\right\rangle \left\langle
n\right\vert \\
S_{+}(p)  &  =\sum_{n=0}^{\infty}(n+1)\left\vert n\right\rangle \left\langle
n+1\right\vert \\
S_{-}(p)  &  =\sum_{n=0}^{\infty}(2p-n)\left\vert n+1\right\rangle
\left\langle n\right\vert .
\end{align*}
Finally, the vector $\left\vert R_{\theta}\right\rangle $ is a coherent state,
parametrized by the twisting angle $\theta$ \cite{NoteTheta}
\begin{equation}
\left\vert R_{\theta}(p)\right\rangle =\sum_{n=0}^{\infty}\frac{(-\cot
\frac{\theta}{2})^{n}(S_{-})^{n}}{n!}\left\vert 0\right\rangle =\sum
_{n=0}^{\infty}\frac{(-\cot\frac{\theta}{2})^{n}}{n!}\binom{2p}{n}\left\vert
n\right\rangle , \label{CoherentState}%
\end{equation}
where $\binom{2p}{n}$ is a generalized binomial coefficient.

Note that the Lindblad operators (\ref{Lindblad Action}) differ from the ones
considered in \cite{MPA2013} by a cyclic permutation $\sigma^{z},\sigma
^{x},\sigma^{y}\rightarrow\sigma^{x},\sigma^{y},\sigma^{z}$, accounted for by
introducing a unitary matrix $U$ which performs a cyclic permutation on the
basis of Pauli matrices,
\begin{equation}
U=\frac{1}{\sqrt{2}}%
\begin{pmatrix}
1 & -i\\
1 & i
\end{pmatrix}
, \label{Umatrix}%
\end{equation}
i.e. $U\sigma^{x}U^{\dagger}=\sigma^{y},U\sigma^{y}U^{\dagger}=\sigma
^{z},U\sigma^{z}U^{\dagger}=\sigma^{x}$.

Some care should be taken in distinguishing the \textit{physical} Hilbert
space $\mathbb{C}^{2}$ in which the matrices $\Omega(p)$ and $U$ are acting,
and an \textit{auxiliary} space $\Re$ spanned by the vectors $\{\left\vert
n\right\rangle \}_{n=0}^{\infty}$, in which $S_{z}(p),S_{\pm}(p)$ are acting.
The result of the operation (\ref{S_N}) $\left\langle 0\right\vert
\Omega^{\otimes_{N}}\left\vert R(\theta)\right\rangle $ is a scalar in the
auxiliary space and is a matrix in the full Hilbert space for $N$ spins $(
\mathbb{C}^{2})^{\otimes N}$. For the following it is convenient to rewrite
the (\ref{RhoNess}) in the form
\begin{equation}
\rho_{NESS}=\left\langle 0,0\right\vert \mathbf{\Omega}(p)^{\otimes_{N}%
}\left\vert R_{\theta},R_{\theta}^{\ast}\right\rangle \label{NESS-OMEGA}%
\end{equation}
where the matrix
\begin{equation}
\label{Omega}\mathbf{\Omega}(p)=U\Omega(p)\otimes_{au}\Omega^{T}%
(-p)U^{\dagger}%
\end{equation}
acts in $\mathbb{C}^{2}\otimes\Re\otimes\Re$ such that $\otimes_{au}$ is a
tensor product in the auxiliary space, and transposition $\Omega^{T}$ is done
in the physical space only. Notation simplifies by embedding the matrices
$\Omega(p)$ and $\Omega^{T}(-p)$ in $\mathbb{C}^{2}\otimes\Re\otimes\Re$
through the redefinitions $\sigma^{a} \to\sigma^{a} \otimes I \otimes I$,
$S_{a} \to I \otimes S_{a} \otimes I$, where the $I$ are the unit operators on
the respective subspaces in the tensor space $\mathbb{C}^{2}\otimes\Re
\otimes\Re$ and by introducing operators $T_{a} \equiv I \otimes I \otimes
T_{a}$ where the operators $T_{z},T_{\pm}$ also satisfy $SU(2)$ and the
representation for $T_{\alpha}(p)$ are obtained from those for $S_{\alpha}(p)$
by the replacement $p\rightarrow-p$. With these definitions one has
$\Omega(p)=S_{z}(p)\sigma^{z}+S_{+}(p)\sigma^{+}+S_{-}(p)\sigma^{-}$,
$\Omega^{T}(-p)=T_{z}(p)\sigma^{z}+T_{+}(p)\sigma^{-}+T_{-}(p)\sigma^{+}$ and
(\ref{Omega}) can be written without the tensor product symbol over the
auxiliary space in the simpler form
\begin{equation}
\mathbf{\Omega}(p)=U\Omega(p)\Omega^{T}(-p)U^{\dagger}. \label{OMEGA(p)}%
\end{equation}

The vector $\left\langle 0,0\right\vert =\left\langle 0\right\vert
\otimes\left\langle 0\right\vert $ in (\ref{NESS-OMEGA}) is a tensor product
of two lowest weight vectors $\left\langle 0,0\right\vert S_{-}=\left\langle
0,0\right\vert T_{-}=0$, and the vector $\left\vert R_{\theta},R_{\theta
}^{\ast}\right\rangle =\left\vert R_{\theta}^{\ast}\right\rangle
\otimes\left\vert R_{\theta}^{\ast}\right\rangle $ is a tensor product of two
coherent states (\ref{CoherentState})%
\begin{equation}
\left\vert R_{\theta},R_{\theta}^{\ast}\right\rangle =\sum_{n,m=0}^{\infty
}\frac{(\cot\frac{\theta}{2})^{n+m}}{n!m!}\binom{2p}{n}\binom{-2p}%
{m}\left\vert n,m\right\rangle . \label{TensorProductCoherentStates}%
\end{equation}

\section{ Steady state magnetization profiles}

\label{sec::Magnetization Profiles}

Steady state expectations are found by the usual prescription, e.g. one-point
correlations are
\begin{equation}
\langle\sigma_{m}^{\alpha}\rangle=\frac{Tr(\sigma_{m}^{\alpha}\rho_{NESS}%
)}{Z(N,\theta)} \label{OnePointExpectations}%
\end{equation}
where $Z(N,\theta)=Tr(\rho_{NESS})$. It is convenient to introduce operators
$B_{\alpha}(p)=Tr(\sigma^{\alpha}\mathbf{\Omega}(p))$,
\begin{align}
B_{0}  &  =2S_{z}T_{z}+S_{+}T_{+}+S_{-}T_{-},\\
B_{x}  &  =S_{+}T_{+}-S_{-}T_{-},\\
B_{y}  &  =S_{z}(T_{-}-T_{+})+T_{z}(S_{-}-S_{+}),\\
B_{z}  &  =i(S_{z}(T_{+}+T_{-})-T_{z}(S_{+}+S_{-})),
\end{align}
which act in the auxiliary space $\Re\otimes\Re$ (here and below we omit the
argument $p$ for simplicity). Note that by construction $[S_{\alpha},T_{\beta
}]=0$ for any $\alpha,\beta$. In terms of the $B$-operators, the normalization
factor becomes
\begin{align}
Z(N,\theta)  &  =Tr(\rho_{NESS})=Tr\left\langle 0,0\right\vert \mathbf{\Omega
}(p)^{\otimes_{N}}\left\vert R_{\theta},R_{\theta}^{\ast}\right\rangle
=\label{l1}\\
&  =\left\langle 0,0\right\vert (Tr(\mathbf{\Omega}(p)))^{_{N}}\left\vert
R_{\theta},R_{\theta}^{\ast}\right\rangle =\label{l2}\\
&  =\left\langle 0,0\right\vert (Tr(\sigma^{0}\mathbf{\Omega}(p))^{_{N}%
}\left\vert R_{\theta},R_{\theta}^{\ast}\right\rangle =\label{l3}\\
&  =\left\langle 0,0\right\vert B_{0}^{N}\left\vert R_{\theta},R_{\theta
}^{\ast}\right\rangle \label{Normalization}%
\end{align}
Analogously, the one-point expectations (\ref{OnePointExpectations}) are
expressed in terms of $B_{\alpha}$ as
\begin{equation}
M_{k,N}^{\alpha}=\langle\sigma_{k}^{\alpha}\rangle=\frac{\left\langle
0,0\right\vert B_{0}^{k-1}B_{\alpha}B_{0}^{N-k})\left\vert R_{\theta
},R_{\theta}^{\ast}\right\rangle }{\left\langle 0,0\right\vert B_{0}%
^{N}\left\vert R_{\theta},R_{\theta}^{\ast}\right\rangle }.
\label{OnePointExpectationsM}%
\end{equation}

Using the $SU(2)$ commutation rules for $S_{\alpha}$ and $T_{\alpha}$, we
obtain, with some effort:
\begin{equation}
\left[  B_{0}\left[  B_{0},B_{x}\right]  \right]  +2\{B_{0},B_{x}%
\}=4(C_{S}+C_{T})B_{x}, \label{B0B0Bz commutation}%
\end{equation}
where $C_{S}$ and $C_{T}$ are $SU(2)$ Casimir operators $C_{S}=S_{z}%
(S_{z}-I)+S_{+}S_{-},C_{T}=T_{z}(T_{z}-I)+T_{+}T_{-}$. From the
representations we readily find $C_{S}=p(p+1),C_{T}=p(p-1)$, so that the
Eq.(\ref{B0B0Bz commutation}) becomes $\left[  B_{0}\left[  B_{0}%
,B_{x}\right]  \right]  +2\{B_{0},B_{x}\}=8p^{2}B_{x}$. Due to the rotational
isotropy of the $XXX$ model, this relation is also valid for other spin
components,
\begin{equation}
\left[  B_{0}\left[  B_{0},B_{\alpha}\right]  \right]  +2\{B_{0},B_{\alpha
}\}=8p^{2}B_{\alpha},\text{ \ \ \ }\alpha=x,y,z. \label{B0B0Bxyz commutation}%
\end{equation}

\begin{figure}[ptbh]
\begin{center}
\subfigure[\label{fig:a}]
{\includegraphics[width=0.4\textwidth]{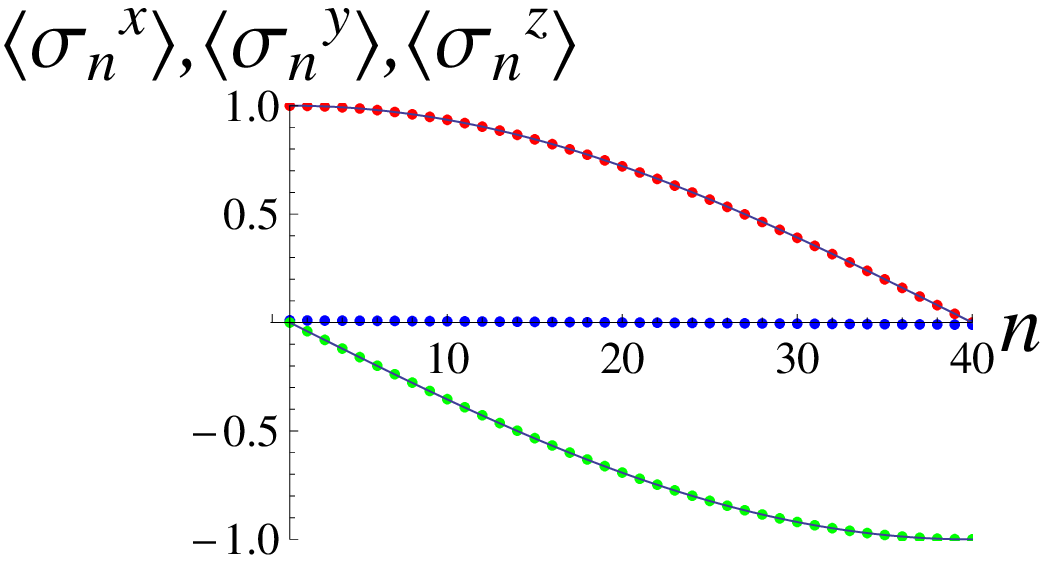}} \qquad
\subfigure[\label{fig:d}]{\includegraphics[width=0.4\textwidth]{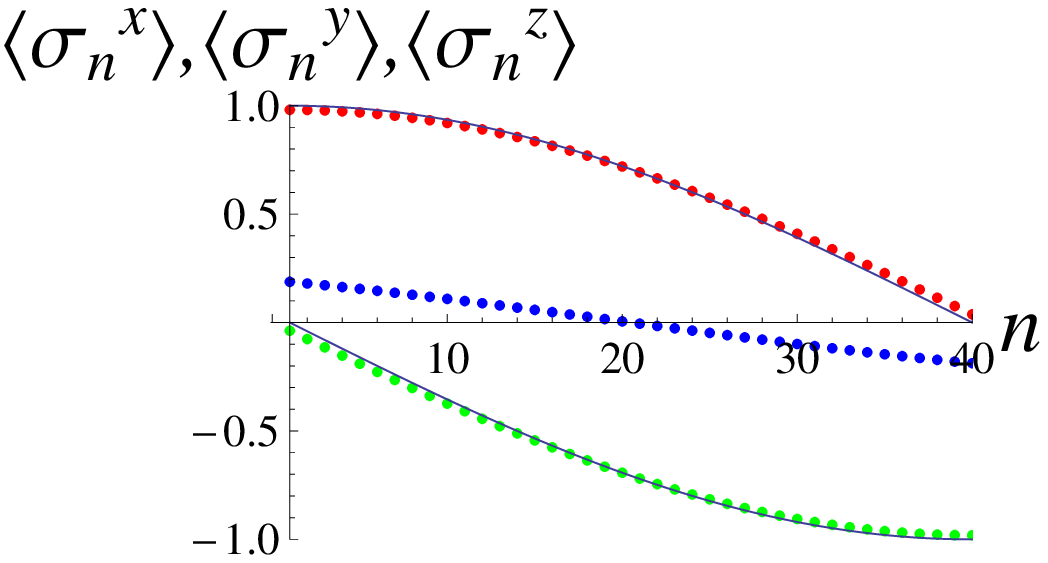}}
\end{center}
\caption{(Color online) Exact steady state magnetization profiles from the MPA
for the $XXX$-chain of $N=40$ sites and $\theta=-\pi/2$. The three components
are $x$ (top), $y$ (bottom) and $z$ (middle). The continuous lines shows
analytic results (\ref{sigmaX}),(\ref{sigmaZ}) valid for large $\Gamma N$.
Panel (a): $\Gamma=4$, Panel (b): $\Gamma=0.2$.}%
\label{FigXYZprofiles}%
\end{figure}

Another important relation is derived by analyzing the behaviour of the
quantity
\begin{equation}
\lim_{\Gamma\rightarrow\infty}\frac{Z(N+1,\theta)}{Z(N,\theta)}=\frac
{4}{\theta^{2}}N^{2}+O(N)\text{ }\label{RatioOfTraces1}%
\end{equation}
\ for $N\gg1$, see Appendix \ref{Appendix::Computation of the Z(N,theta)}.
Multiplying (\ref{B0B0Bxyz commutation}) by $\left\langle 0,0\right\vert
B_{0}^{k-1}$ from the left and by $B_{0}^{N-k-1})\left\vert R_{\theta
},R_{\theta}^{\ast}\right\rangle $ from the right, and using
(\ref{RatioOfTraces1}), (\ref{OnePointExpectationsM}), we obtain%
\begin{align}
(-2M_{k+1,N+1}^{\alpha}+M_{k+2,N+1}^{\alpha}+M_{k,N+1}^{\alpha})\frac
{Z(N+1,\theta)}{Z(N,\theta)} &  \nonumber\\
+2(M_{k,N}^{\alpha}+M_{k+1,N}^{\alpha})=8p^{2}M_{k,N-1}^{\alpha}%
\frac{Z(N-1,\theta)}{Z(N,\theta)} &  .\label{DiscreteEqForM}%
\end{align}
Taking the continuum limit, we substitute $k/N=x,$ $M_{k,N}^{\alpha}%
,M_{k+1,N}^{\alpha}\rightarrow M_{\alpha}(x),M_{\alpha}(x+\frac{1}{N})$ in the
above and expand in Taylor series in $1/N$. In the lowest order of expansion,
we can neglect the right-hand side of (\ref{DiscreteEqForM}) and obtain using
(\ref{RatioOfTraces1}),
\begin{equation}
\frac{\partial^{2}M^{\alpha}(x)}{\partial x^{2}}+\theta^{2}M^{\alpha
}(x)=0,\label{DiffEqForM}%
\end{equation}
provided that
\begin{equation}
\Gamma\gg\Gamma^{\ast}=\frac{1}{N},\label{GammaTypicalValue}%
\end{equation}
Integrating (\ref{DiffEqForM}) with the boundary conditions $M^{\alpha
}(0)=\sigma_{\text{target(L)}}^{a}$, $M^{\alpha}(1)=\sigma_{\text{target(R)}%
}^{\alpha}$ where $\sigma_{\text{target(L,R)}}^{a}$ are targeted boundary
magnetizations,
\begin{equation}
\sigma_{\text{target(L)}}^{x}=1;\sigma_{\text{target(L)}}^{y}=\text{ }%
\sigma_{\text{target(L)}}^{z}=0.\label{TargetedLeft}%
\end{equation}%
\begin{equation}
\sigma_{\text{target(R)}}^{x}=\cos\theta;\text{ }\sigma_{\text{target(R)}}%
^{y}=\sin\theta;\text{ }\sigma_{\text{target(R)}}^{z}=0,\label{TargetedRight}%
\end{equation}
we obtain stationary density profiles
\begin{align}
M^{x}(x) &  =\cos\theta x\nonumber\\
M^{y}(x) &  =\sin\theta x\label{MagnetizProfilesContinuous2}\\
M^{z}(x) &  =0,\nonumber
\end{align}
interpolating between the left and right boundary values, see
Fig.\ref{FigXYZprofiles}. In a finite chain these asymptotic results can be
approximated by
\begin{align}
\langle\sigma_{k}^{x}\rangle &  =\cos\left(  \theta\frac{k-1}{N-1}\right)
,\label{sigmaX}\\
\langle\sigma_{k}^{y}\rangle &  =\sin\left(  \theta\frac{k-1}{N-1}\right)
,\label{sigmaZ}%
\end{align}
see Fig.\ref{FigXYZprofiles}. Note that for $\theta=\pi$ the
Eqs(\ref{DiffEqForM},\ref{MagnetizProfilesContinuous2}) reproduce the results
obtained in \cite{ProsenExact2011}.

Numerical evidence as well as the leading behaviour of the normalization
(\ref{RatioOfTraces1}) suggest that $\theta^{2}/N$ is a good scaling variable.
For small $\Gamma$ of order of $\theta^{2}/N$, there are corrections to the
asymptotic formula (\ref{MagnetizProfilesContinuous2}). In particular the
$M^{z}(x)$ profile becomes harmonic as well $M^{z}(x)=-f_{1}(\Gamma
)\sin((x-\frac{1}{2})\omega(\Gamma))$ see Fig.\ref{FigXYZprofiles}(b).

It is quite remarkable that the stationary magnetization profile for NESS
satisfies a simple harmonic equation (\ref{DiffEqForM}) for all magnetization
components. However we were not able to derive the (\ref{DiffEqForM}) in a
shorter way. Note that the NESS is formed with highly excited states, since
the magnetization is not alternating in space as would be expected for a
ground state of a antiferromagnet (\ref{Hamiltonian}). Reversing the sign of
the spin exchange interaction $H\rightarrow-H$ amounts, for the NESS, to
reversing the sign $\Gamma$, see (\ref{LME}). We find that $M^{x}(x)$ and
$M^{y}(x)$ are invariant under $\Gamma\rightarrow-\Gamma$ exchange, while
$M^{z}(x)$ reverses its sign, $M^{z}(x)\rightarrow-M^{z}(x)$. As a
consequence, the limiting harmonic shape of the magnetization profile for
$\Gamma\gg\Gamma^{\ast}=\frac{1}{N}$, given by
(\ref{MagnetizProfilesContinuous2}), does not depend on whether the bulk
Hamiltonian is antiferromagnetic or ferromagnetic one, as long as it stays
isotropic \cite{AnisotropicMagnetizationProfiles}. Note also that harmonic
density profile leads to diffusive $1/N$ scaling of the trasversal
magnetization current, see discussion after (\ref{Jz(N)limit}).

%%%%/////////////// Two Figures //////////////////////////////

\section{Magnetization currents}

\label{sec::Magnetization Currents}

The $XXX$-model, due to its isotropy, has three independent local
magnetization currents defined by the time derivative of the respective local
magnetization components, $\frac{d\sigma_{n}^{\alpha}}{dt}=\hat{\jmath}%
_{n-1}^{\alpha}-\hat{\jmath}_{n}^{\alpha}$, where
\begin{equation}
\hat{\jmath}_{n}^{\alpha}=2\varepsilon_{\alpha\beta\gamma}\sigma_{n}^{\beta
}\sigma_{n+1}^{\gamma}, \label{Jn^alpha}%
\end{equation}
with the Levi-Civita symbol $\varepsilon_{\alpha\beta\gamma}$. In the steady
state the current expectations $j^{\alpha}(\theta):= \langle\hat{\jmath}%
_{n}^{\alpha}\rangle$ are position-independent.

Some important conclusions can be drawn on the base of LME symmetries, and the
uniqueness of the steady state (\ref{RhoNess}). Parallel boundary driving
$\theta=0$ does not create any gradient, and therefore all currents vanish,
$j^{\alpha}(0)=0$. For the antiparallel alignment $\theta=\pi$, the
magnetization currents $j^{z},j^{y}$ vanish, but not $j^{x}$, $j^{z}%
(\pi)=j^{y}(\pi)=0$. To see this, note that for $\theta=\pi$ the NESS has a
symmetry $\rho_{NESS}=\Sigma_{x}\rho_{NESS}\Sigma_{x}$, where $\Sigma
_{x}=(\sigma_{x})^{\otimes N}$. The operators $\hat{\jmath}_{n}^{y}%
,\hat{\jmath}_{n}^{z}$ change sign under the action of the\ symmetry :
$\Sigma_{x}\hat{\jmath}_{n}^{y}\Sigma_{x}=-\hat{\jmath}_{n}^{y},$
$\ \Sigma_{x}\hat{\jmath}_{n}^{z}\Sigma_{x}=-\hat{\jmath}_{n}^{z}$. Since
$Tr(\hat{\jmath}_{n}^{y,z}\rho_{NESS})=-Tr(\Sigma_{x}\hat{\jmath}_{n}%
^{y,z}\Sigma_{x}\rho_{NESS})=-Tr(\hat{\jmath}_{n}^{y,z}\Sigma_{x}\rho
_{NESS}\Sigma_{x})=-Tr(\hat{\jmath}_{n}^{y,z}\rho_{NESS})$, the current
suppression follows, see \cite{conLivi,XYtwist} for more details.

In terms of the MPA the steady magnetization currents are given by
\begin{equation}
\langle\hat{\jmath}_{k}^{\alpha}\rangle=\frac{2\varepsilon_{\alpha\beta\gamma
}\left\langle 0,0\right\vert B_{0}^{k-1}B_{\beta}B_{\gamma}B_{0}%
^{N-k-1})\left\vert R_{\theta},R_{\theta}^{\ast}\right\rangle }{Z(N,\theta)}.
\label{MagnetizCurrentExpectation}%
\end{equation}
Using the algebra, we find%
\begin{align}
\lbrack B_{y},B_{z}]  &  =2i(T_{z}-S_{z})B_{0}\label{CommBxBy}\\
\lbrack B_{x},B_{y}]  &  =B_{0}I_{1}=B_{0}(S_{+}-S_{-}+T_{+}-T_{-}%
)\label{CommBzBx}\\
i[B_{z},B_{x}]  &  =B_{0}I_{2}=B_{0}(S_{+}+S_{-}-T_{+}-T_{-}) \label{CommByBz}%
\end{align}

\begin{figure}[ptbh]
\begin{center}
\subfigure[\label{figX:d}]
{\includegraphics[width=0.4\textwidth]{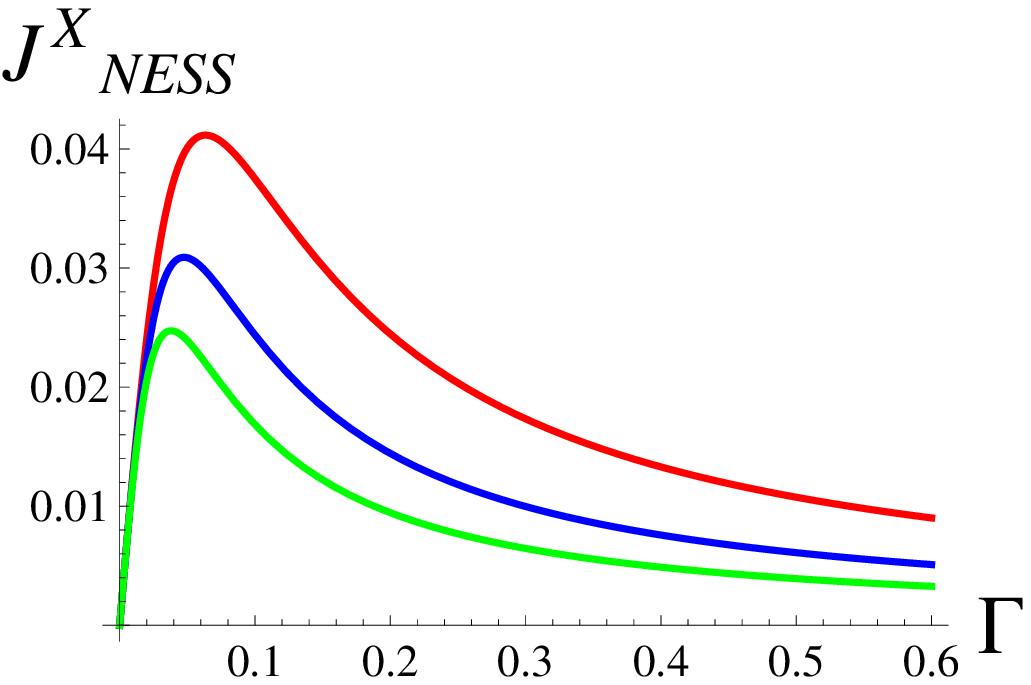}} \qquad
\subfigure[\label{figX:d}]
{\includegraphics[width=0.4\textwidth]{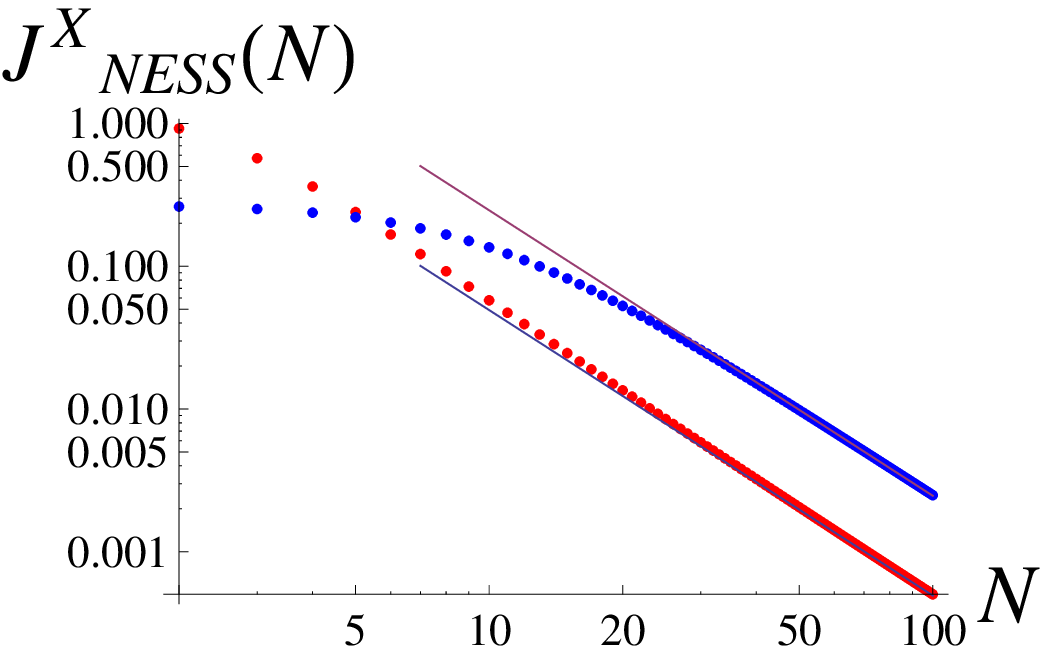}}
\end{center}
\caption{(Color online) Exact $j^{x}$ steady state currents from MPA
(\ref{Jx(N)}), for $\theta=\pi/2$, as function of $\Gamma$ for $N=30,40,50$%
(upper,middle and lower curve) (Panel(a)), and as function of system size for
two fixed values of $\Gamma=1,0.2 $(lower and upper curve) (Panel(b)). Lines
on Panel (b) show the asymptotics (\ref{Jx(N)limit}).}%
\label{FigXcurrent}%
\end{figure}

Operators $(T_{z}-S_{z}),I_{1},I_{2}$ commute with $B_{0}$, which manifests
the local conservation of the magnetizations $\sigma_{n}^{z},\sigma_{n}%
^{y},\sigma_{n}^{x}$ respectively, so that current expectations values
$\langle\hat{\jmath}_{n}^{\alpha}\rangle=j^{\alpha}$ are indeed
position--independent. We readily derive the exact expressions for the $x$ --
magnetization current as
\begin{equation}
j^{x}(N)=-8ip\frac{Z(N-1,\theta)}{Z(N,\theta)}, \label{Jx(N)}%
\end{equation}
which has a characteristic bell-shaped form, see Fig.\ref{FigXcurrent}.

For $N=2,3,4$ the exact expressions for $j^{x}$ are
\begin{equation}
j^{x}(2)=\frac{16\Gamma\sin^{2}\left(  \frac{\theta}{2}\right)  (3+\cos
\theta)}{8\ \Gamma^{2}+19+12\cos\theta+\cos2\theta},
\end{equation}

\begin{equation}
j^{x}(3)=\frac{16\Gamma\sin^{2}\left(  \frac{\theta}{2}\right)  (8\Gamma
^{2}+19+12\ \cos\theta+\cos2\theta)}{48\Gamma^{4}+208\Gamma\ ^{2}+126+\left(
16\Gamma^{4}+112\Gamma^{2}+111\right)  \cos\theta+18\cos2\theta+\cos3\theta}%
\end{equation}

\begin{equation}
j^{x}(4)=\frac{16\Gamma\sin^{2}\left(  \frac{\theta}{2}\right)  \left(
48\Gamma^{4}+208\ \Gamma^{2}+126+\left(  16\Gamma^{4}+112\Gamma^{2}%
+111\right)  \cos\theta+18\cos\ 2\theta+\cos3\theta\right)  }{\gamma
_{0}+\gamma_{1}\cos\ \theta+\gamma_{2}\cos2\theta+24\cos3\theta+\cos4\theta}%
\end{equation}
with $\gamma_{0}=688\Gamma^{6}+3584\Gamma^{4}+3616\ \Gamma^{2}+867$,
$\gamma_{1}=8\left(  56\Gamma^{6}+320\Gamma^{4}+400\Gamma^{2}+117\right)  $,
$\gamma_{2}=4\left(  4\Gamma^{6}+32\Gamma^{4}+88\Gamma^{2}+55\right)  $.

For sufficiently large $\Gamma$,$N$ using (\ref{RatioOfTraces1}), we obtain
\begin{equation}
j^{x}(N)=\frac{2}{\Gamma}\frac{\theta^{2}}{N^{2}}+O\left(  N^{-3}\right)
,\label{Jx(N)limit}%
\end{equation}
Note that the applicability of the asymptotic formula (\ref{Jx(N)limit}), as
well as of other asymptotics (\ref{Jz(N)limit}),(\ref{RatioOfTraces1}%
),(\ref{sigmaX}), (\ref{sigmaZ}) depends on the value of the coupling $\Gamma$
which should be much larger than a characteristic value $\Gamma^{\ast}$, see
(\ref{GammaTypicalValue}). For small values of $\Gamma\ll\Gamma^{\ast}$, we
find linear growth of $j^{x}$ with $\Gamma$ of the form%
\[
\left.  j^{x}(N)\right\vert _{\Gamma\ll\Gamma^{\ast}}=\frac{8\sin^{2}%
\frac{\theta}{2}}{3+\cos\theta}\Gamma+O\left(  \Gamma^{2}\right)  ,
\]
valid for all $N\geq2$. Finally, for intermediate $\Gamma$ values,
$\Gamma=O(\Gamma^{\ast})$ the function $j^{x}$ has a maximum in $\Gamma$, see
Fig.\ref{FigXcurrent}, which scales as $1/N$. The value of $\Gamma$ which
maximizes $j^{x}$ is well--aproximated by $\Gamma_{\max}=2/N$, for $N\gg1$,
data not shown. The amplitude of the maximum can be approximated as $\left.
j^{x}(N)\right\vert _{\Gamma=\Gamma_{\max}}=\theta^{2}/(2N)$.

The $j^{y}(N)$ current has the same scaling $1/N^{2}$ for $\Gamma\gg
\Gamma^{\ast}$ as the current $j^{x}$ as it is simply proportional to the
$j^{x}$ current, with a proportionality coefficient $(-\cot\frac{\theta}{2})$.
Indeed, e.g. for $\theta=\pm\pi/2$ it follows from the properties of the
coherent state that $I_{2}\left\vert R_{\theta},R_{\theta}^{\ast}\right\rangle
=\mp4p\left\vert R_{\theta},R_{\theta}^{\ast}\right\rangle $, and we obtain
from (\ref{CommByBz}), (\ref{MagnetizCurrentExpectation}) that $j^{x}%
(\theta=\pm\pi/2)=\mp j^{z}(\theta=\pm\pi/2)$. For arbitrary $\theta$,
$I_{2}\left\vert R_{\theta},R_{\theta}^{\ast}\right\rangle =(2p(\cot
\frac{\theta}{2}+\cot^{-1}\frac{\theta}{2})I+(\cot\frac{\theta}{2}-\cot
^{-1}\frac{\theta}{2})(S-T))\left\vert R_{\theta},R_{\theta}^{\ast
}\right\rangle $ and after a straightforward calculation we obtain
\begin{equation}
j^{y}(N,\theta)=8ip\frac{Z(N-1,\theta)}{Z(N,\theta)}\cot\frac{\theta}{2}%
=-\cot\frac{\theta}{2}\times j^{x}(N,\theta). \label{Jy}%
\end{equation}

\begin{figure}[ptbh]
\begin{center}
\subfigure[\label{figX:d}]
{\includegraphics[width=0.4\textwidth]{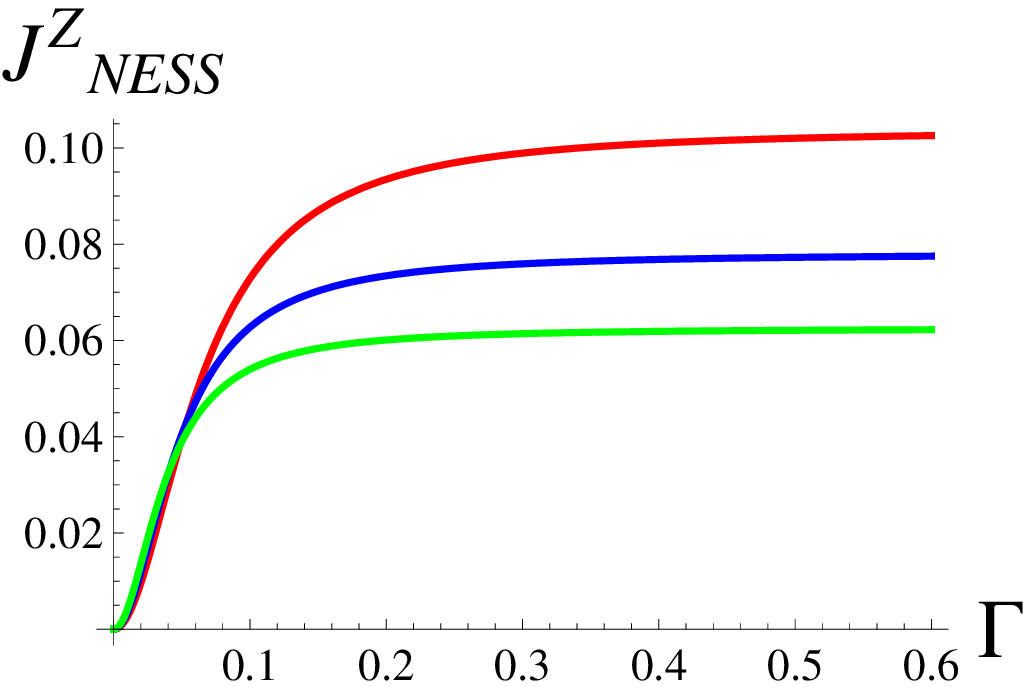}} \qquad
\subfigure[\label{figX:d}]
{\includegraphics[width=0.4\textwidth]{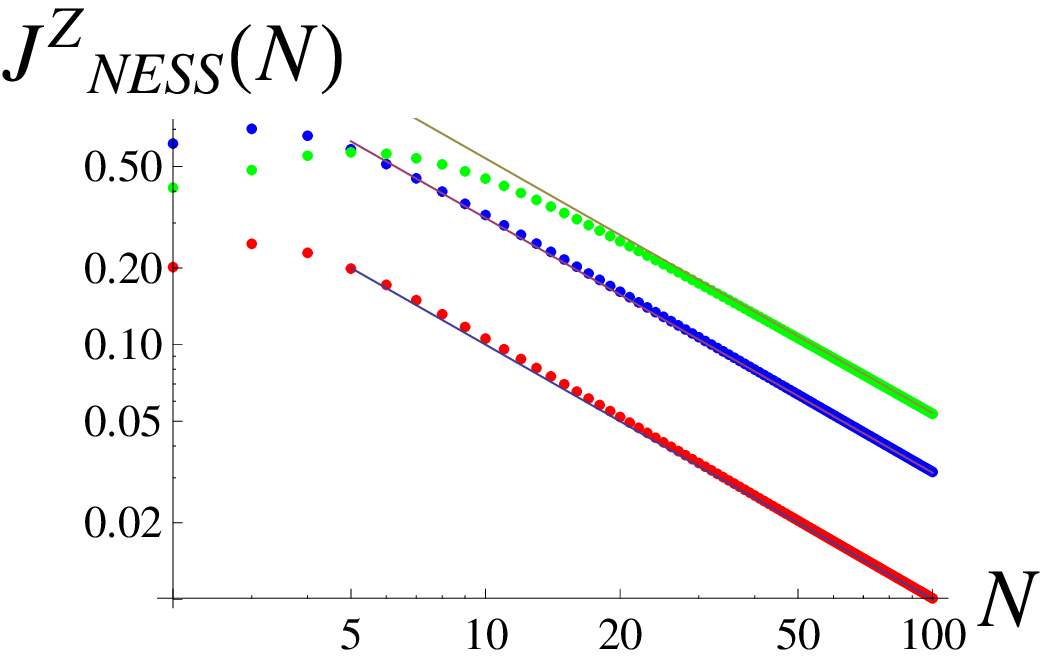}}
\end{center}
\caption{(Color online) Exact $j^{z}$ steady state current from MPA, for
$\theta=\pi/2$, as function of $\Gamma$ for $N=30,40,50$(upper,middle and
lower curve) (Panel(a)) and as function of system size for fixed $\Gamma=1$
and various $\theta=0.5,\pi/2,6\pi/7$ (lower, middle and upper curve)
(Panel(b)). Lines on Panel (b) show the asymptotics (\ref{Jz(N)limit}).}%
\label{FigZcurrent}%
\end{figure}

On the contrary, the scaling and qualitative behaviour of the $j^{z}$ current
is different, in two respects: It scales as $1/N$ and is a monotonically
increasing function of $\Gamma$, see Fig. \ref{FigZcurrent}. Using properties
of coherent states one obtains the following exact expression%
\begin{equation}
j^{z}(N,\theta)=\frac{4}{\sin\theta}\frac{\left\langle 0,0\right\vert
B_{0}^{N-1}(-S_{z}-T_{z}))\left\vert R_{\theta},R_{\theta}^{\ast}\right\rangle
}{Z(N,\theta)} \label{Jz}%
\end{equation}

For $N=2,3,4$ explicit expressions for $j^{z}(N)$ are%
\begin{equation}
j^{z}(2)=\frac{16\Gamma^{2}\sin\theta}{8\Gamma^{2}+19+12\cos\theta+\cos
2\theta}%
\end{equation}

\begin{equation}
j^{z}(3)=\frac{64\Gamma^{2}\sin\theta\left(  \Gamma^{2}+3+\cos\theta\right)
}{48\ \Gamma^{4}+208\Gamma^{2}+126+\left(  16\Gamma^{4}+112\Gamma
^{2}+111\right)  \cos\ \theta+18\cos2\theta+\cos3\theta}%
\end{equation}

\begin{equation}
j^{z}(4)=\frac{32\Gamma^{2}\sin\theta\left(  20\Gamma\ ^{4}+92\Gamma
^{2}+57+(4\Gamma^{4}+28\Gamma^{2}+36)\cos\theta+3\cos2\theta\right)  }%
{\beta_{0}+\beta_{1}\cos\theta+\beta_{2}\cos2\theta+24\cos3\theta
\ +\cos4\theta}%
\end{equation}
where $\beta_{0}=688\Gamma^{6}+3584\Gamma^{4}+3616\Gamma^{2}+867$, $\beta
_{1}=448\Gamma^{6}+2560\Gamma^{4}+3200\Gamma^{2}+936$, $\beta_{2}=\Gamma
^{6}+128\Gamma^{4}+352\Gamma^{2}+220$. For $\theta=\pi/2$, the above
expressions reduce to those obtained by direct diagonalization in
\cite{XYtwist}, $j^{z}(2)=\frac{8\Gamma^{2}}{4\Gamma^{2}+9}$, $j^{z}%
(3)=\frac{16\Gamma^{2}\left(  \Gamma^{2}+3\right)  }{12\Gamma^{4}+52\Gamma
^{2}+27}$, $j^{z}(4)=\frac{8\Gamma^{2}\left(  10\Gamma^{4}+46\Gamma
^{2}+27\right)  }{3\left(  28\Gamma^{6}+144\Gamma^{4}+136\Gamma^{2}+27\right)
}$. For small values of $\Gamma\ll\Gamma^{\ast}$, we find quadratic growth of
$j^{z}$ with $\Gamma$ of the form%
\begin{equation}
\left.  j^{z}(N)\right\vert _{\Gamma\ll\Gamma^{\ast}}=\frac{8\sin\theta
}{(3+\cos\theta)^{2}}(N-1)\Gamma^{2}+O\left(  \Gamma^{4}\right)  ,
\label{JzForSmallGamma}%
\end{equation}
for all $N\geq2$. So,

For $\theta=\pi/2$ Eq(\ref{JzForSmallGamma}) reduces to obtained in \cite{Weak
XXZ} by direct diagonalization. Finally, for large $\Gamma,N$, (\ref{Jz})
reduces to a very simple expression,
\begin{equation}
\left.  j^{z}(N)\right\vert _{\Gamma\gg\Gamma^{\ast},N\gg1}=\frac{2\theta}%
{N}+O\left(  \frac{1}{N^{2}}\right)  \label{Jz(N)limit}%
\end{equation}
see Appendix \ref{Appendix::Computation of the Z(N,theta)}. Note that unlike
the currents $j^{x},j^{y}$ which decrease with the coupling as $1/\Gamma$, the
current $j^{z}$ \textit{increases} with $\Gamma$ and takes a \textit{finite}
value (\ref{Jz(N)limit}) as $\Gamma\rightarrow\infty$, see also Fig.
\ref{FigZcurrent}. The finiteness of the magnetization current $j^{z}$ at
$\Gamma\rightarrow\infty$ also emerges from an analysis of a perturbative
expansion of NESS for small systems in $1/\Gamma$, for arbitrary spin exchange
$Z-$anisotropy $\Delta\neq0$ \cite{XYtwist}. Note also that an increase of
system size $N$ results in increase of $j^{z}(N)$ for small coupling
$\Gamma\ll\Gamma^{\ast}$due to (\ref{JzForSmallGamma}), and in decrease of
$j^{z}(N)$ for $\Gamma\gg\Gamma^{\ast}$ due to (\ref{Jz(N)limit}).

Interestingly, the leading term in the asymptotic expression (\ref{Jz(N)limit}%
) for the current $j^{z}$ can be obtained from the exact density profiles
(\ref{MagnetizProfilesContinuous2}) if one neglects the connected part of the
two-point correlations, as
\begin{equation}
j_{MF}^{z}(N)=2\langle\sigma_{n}^{x}\rangle\langle\sigma_{n+1}^{y}%
\rangle-2\langle\sigma_{n}^{y}\rangle\langle\sigma_{n+1}^{x}\rangle.
\label{JzMF}%
\end{equation}
Making use of the exact expressions (\ref{MagnetizProfilesContinuous2}), we
have $\langle\sigma_{n}^{x}\rangle=\cos\left(  \theta\frac{n}{N}\right)  $,
$\langle\sigma_{n+1}^{y}\rangle=\sin\left(  \theta\frac{n+1}{N}\right)  $. The
Eq.(\ref{Jz(N)limit}) is then obtained in the first non-vanishing order of the
Taylor expansion of (\ref{JzMF}) in $\frac{1}{N}$.

\begin{figure}[ptbh]
\centerline{\includegraphics[width=0.4\textwidth]{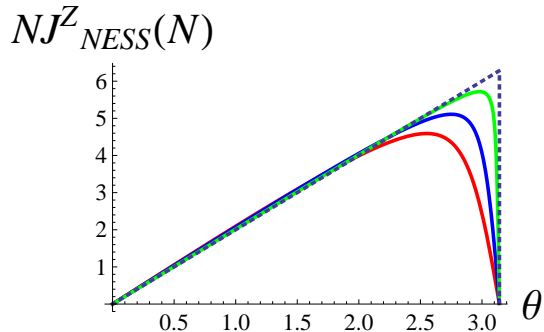}}
\caption{(Color online) Function $Nj^{z}(N)$ versus $\theta$, for
$N=10,20,100$ (lower, middle, upper curves). In all cases, ${\frac{\theta^{2}
}{{\Gamma N}}} \ll1$. The dotted line shows the asymptotics (\ref{Jz(N)limit})
for $N\rightarrow\infty$. }%
\label{FigThetaJzScaling}%
\end{figure}

Another surprising property consists in the fact that for large $N$ the
maximum of the $j^{z}(N,\theta)$ current is observed for $\theta=\pi
-O(N^{-1})$. On the other hand, see the discussion after (\ref{Jn^alpha}),
$j^{z}(N,\pi)=0$ for any $N$. Consequently, the limits $N\rightarrow\infty$
and $\theta\rightarrow\pi$ do not commute, namely
\begin{align}
\lim_{\theta\rightarrow\pi}\lim_{N\rightarrow\infty}Nj^{z}(N)  &  =2\pi\\
\lim_{N\rightarrow\infty}\lim_{\theta\rightarrow\pi}Nj^{z}(N)  &  =0,
\end{align}
see Fig.\ref{FigThetaJzScaling}. The reason for the non-commutativity of the
limits is the presence of an additional symmetry at $\theta=\pi$ as discussed
after (\ref{Jn^alpha}).

\section{Conclusions}

We have investigated open driven $XXX$ model where boundary spins are pumped
in two different directions, with an arbitrary twisting angle between them. We
find explicit expressions for one- and two-point observables (magnetization
currents and magnetization profiles) in the steady state, and investigate
various asymptotic regimes. We find scaling of the magnetization current to be
qualitatively different in the direction parallel to the twisting plane, and
in the orthogonal direction. We find explicit dependencies on the twisting
angle, and retrieve known cases. At the point of antiparallel driving
$\theta=\pi$ we find a non-analyticity of the transversal magnetization
current in the thermodynamic limit.

It is instructive to compare our findings with previous results on spin
transport in the isotropic Heisenberg chain. For $\theta=\pi$, we retrieve an
anomalous scaling of the current $1/N^{2}$ (\ref{Jx(N)limit}), obtained in
\cite{ProsenExact2011} (note that our $j^{x}$ current then corresponds to the
$j^{z}$ current of \cite{ProsenExact2011}). For $\theta=\pi$, and small
driving (i.e. small amplitude of the targeted boundary values, another scaling
form was obtained, $j=const/\sqrt{N}$ \cite{ZnidaricHeisIsotr2011}. In both
cases the current decreases as $1/\Gamma$ with coupling which was attributed
to a quantum Zeno effect.

Our study shows that already by an infinitesimal perturbation of the
antiparallel boundary alignment $\theta\rightarrow\pi-\varepsilon$ in the
isotropic model an additional current appears, pointing in the direction
perpendicular to the twisting plane, which has yet another scaling form $1/N$
(\ref{Jz(N)limit}). This new current does not decrease with coupling, but, on
the contrary, saturates to its maximal value at $\Gamma\rightarrow\infty$.

\section*{Acknowledgements}

V.P. thanks T. Prosen for a discussion and acknowledges a partial DFG support.

\appendix

\section{Computation of the $Z(N,\theta)$}

\label{Appendix::Computation of the Z(N,theta)}

$Z(N,\theta)$ is given by Eq.(\ref{Normalization}). The operator $B_{0}$,
restricted to the basis of vectors $V=\{\left\vert n,n\right\rangle
\}_{n=0}^{\infty}$, has the form
\[
\left.  B_{0}\right\vert _{V}=\sum_{n=0}^{\infty}2s_{n}t_{n}\left\vert
n,n\right\rangle \left\langle n,n\right\vert +s_{n+1}^{+}t_{n+1}^{+}\left\vert
n,n\right\rangle \left\langle n+1,n+1\right\vert +s_{n}^{-}t_{n}^{-}\left\vert
n+1,n+1\right\rangle \left\langle n,n\right\vert ,
\]
where $s_{n}=p-n;$ $s_{n}^{+}=n\,;s_{n}^{-}=n-2p$ and expressions for
$t_{n},t_{n}^{\pm}$ are obtained from $s_{n},s_{n}^{\pm}$ by replacing
$p\rightarrow p^{\ast}=-p$. Let us introduce functions $F_{n}^{(N)}(p)$
defined as
\begin{equation}
\left\langle 0,0\right\vert B_{0}^{N}=\sum_{n=0}^{N}F_{n}^{(N)}(p)\left\langle
n,n\right\vert \label{F(N) definition}%
\end{equation}
through which the normalization factor is expressed as $Z(N,\theta)=\sum
_{n=0}^{\infty}F_{n}^{(N)}(p)\left\langle n,n\right\vert R_{\theta},R_{\theta
}^{\ast}\rangle$, or,
\begin{equation}
Z(N,\theta)=\sum_{n=0}^{N}F_{n}^{(N)}(p)\left(  \cot\frac{\theta}{2}\right)
^{2n}\binom{2p}{n}\binom{-2p}{n} \label{Z(N) general}%
\end{equation}
\textbf{Limit of small }$p\rightarrow0$\textbf{.} \ For $p\rightarrow0$, in
the lowest order in $p$ we obtain
\begin{equation}
Z(N+1,\theta)=-4p^{2}\left(  F_{1}^{(N)}(0)+\sum_{n=1}^{N+1}F_{n}%
^{(N+1)}(0)\left(  \cot\frac{\theta}{2}\right)  ^{2n}\frac{1}{n^{2}}\right)
+o(p^{2}). \label{Z(N+1)}%
\end{equation}
On the other hand, from the (\ref{F(N) definition}) the recursion relations
for $F_{n}^{(N+1)}(0)$ read
\begin{equation}
F_{n}^{(N+1)}(0)=n^{2}\left(  2F_{n}^{(N)}(0)+F_{n-1}^{(N)}(0)+F_{n+1}%
^{(N)}(0)\right)  \label{RecurrenceRelations}%
\end{equation}
with the initial condition $F_{n}^{(1)}(0)=\delta_{n,1}$. Substituting the
recursion in the (\ref{Z(N+1)}), and noting that $F_{k}^{(N)}=0$ for $k>N$, we
obtain
\begin{equation}
\lim_{p\rightarrow0}\frac{Z(N+1,\theta)}{p^{2}}=-\frac{16}{\sin^{2}\theta}%
\sum_{n=1}^{N}F_{n}^{(N)}(0)\left(  \cot\frac{\theta}{2}\right)  ^{2n}
\label{Z(N)limitSmallP}%
\end{equation}

Here below we treat the case $\theta=\pi/2$ in more detail, for which the
quantity $Z(N+1,\theta)$ becomes%
\begin{equation}
\lim_{p\rightarrow0}\frac{Z(N+1,\pi/2)}{p^{2}}=-16\sum_{n=1}^{N}F_{n}^{(N)}(0)
\label{Z(N,Pi/2)}%
\end{equation}
and the recursion relations for $F_{n}^{(N)}(0)$ are given by
(\ref{RecurrenceRelations}). For further analysis, fix notation as follows:
(i) $F_{n}^{(N)}:=F_{n}^{(N)}(0)$, (ii) $F^{(N)}=\sum_{n=0}^{N}F_{n}^{N}$ The
conjecture is $\lim_{N\rightarrow\infty}1/N^{y}F^{(N+1)}/F^{(N)}=c^{-2}$ with
$y=2$ and $c=4/\pi$ This implies that asymptotically $F^{(N)}=F(cN)^{2N}$ with
some undetermined constant $F$. Solving the recursion explicitly for $n$ close
to $N$ yields the following exact expressions:
\begin{align}
F_{N+1}^{(N)}  &  =0,\\
F_{N}^{(N)}/(N!)^{2}  &  =1,\\
F_{N-1}^{(N)}/[(N-1)!]^{2}  &  =2\sum_{k=1}^{N-1}k^{2}\sim\frac{2N^{3}}{3},\\
F_{N-2}^{(N)}/[(N-2)!]^{2}  &  =4\sum_{k=1}^{N-2}k^{2}\sum_{m=1}^{k}m^{2}%
+\sum_{k=1}^{N-2}k^{2}(k+1)^{2}\sim\frac{1}{2!}\left(  \frac{2N^{3}}%
{3}\right)  ^{2},\\
F_{N-3}^{(N)}/[(N-3)!]^{2}  &  =8\sum_{k=1}^{N-3}k^{2}\sum_{m=1}^{k}m^{2}%
\sum_{n=1}^{k}n^{2}+2\sum_{k=1}^{N-3}k^{2}\sum_{m=1}^{k}m^{2}(m+1)^{2}%
\nonumber\\
&  +2\sum_{k=1}^{N-3}k^{2}(k+1)^{2}\sum_{m=1}^{k+1}m^{2}\sim\frac{1}%
{3!}\left(  \frac{2N^{3}}{3}\right)  ^{3}.
\end{align}
Here $\sim$ means up to order $1/N$. Hence for \textit{fixed} $k$ we obtain
the large $N$ behaviour
\begin{equation}
\lim_{N\rightarrow\infty}N^{-k^{2}}F_{N-k}^{(N)}/[(N-k)!]^{2}=\frac{2^{k}%
}{3^{k}k!}%
\end{equation}

However, these terms grow strongly with $k$ and do not dominate the sum over
$m$. To study the behaviour for $m\propto N$ we make the the ansatz
\begin{equation}
F_{m}^{(N)}/F^{(N)}=\frac{1}{\sqrt{4\pi bN}}\mathrm{e}^{-\frac{(m-aN)^{2}%
}{2bN}} \label{AnsatzForF}%
\end{equation}
This solves the recursion (\ref{RecurrenceRelations}) with $a=c/2$,
$b=c^{2}/8$ and any $m-aN$ of the order $\sqrt{N}$ solves the recursion
(\ref{RecurrenceRelations}) up to corrections of order $1/N$ for any
(positive) value of $c$. This can be seen by setting $m=aN+x\sqrt{N}$ and
plugging this into the recursion (\ref{RecurrenceRelations}).

The l.h.s. yields
\begin{align}
F^{(N+1)}_{m}  &  = F^{(N+1)} \frac{1}{\sqrt{4\pi b (N+1)}} \mathrm{e}%
^{-\frac{(m-a(N+1))^{2}}{2b(N+1)}}\nonumber\\
&  = F^{(N+1)} \frac{1}{\sqrt{1+N^{-1}} } \frac{1}{\sqrt{4\pi b N}} \,
\mathrm{e}^{-\frac{(m-aN)^{2}-2a(m-aN)+a^{2}}{2bN}\frac{1}{1+N^{-1}}%
}\nonumber\\
&  = F^{(N+1)} \frac{1}{\sqrt{1+N^{-1}} } \frac{1}{\sqrt{4\pi b N}} \,
\mathrm{e}^{-\frac{x^{2}-2ax\sqrt{N}+a^{2}}{2bN} \frac{1}{1+N^{-1}}
}\nonumber\\
&  = F^{(N+1)} \left(  1+O(N^{-1})\right)  \frac{1}{\sqrt{4\pi b N}} \,
\mathrm{e}^{-\frac{x^{2}}{2b} + \frac{2ax}{b\sqrt{N}}+O(N^{-1}) }\nonumber\\
&  = F^{(N)}_{m} \frac{F^{(N+1)}}{F^{(N)}}\left(  1+\frac{ax}{b\sqrt{N}%
}+O(N^{-1})\right) \nonumber
\end{align}
On the r.h.s. one gets for $G^{(N)}_{m} := m^{2} (F^{(N)}_{m+1} +
F^{(N)}_{m-1} +2 F^{(N)}_{m}) $%

\begin{align}
G_{m}^{(N)})  &  =F_{m}^{(N)}m^{2}\left[  2+2\mathrm{e}^{\frac{1}{2bN}}%
\cosh{\left(  \frac{m-aN}{bN}\right)  }\right] \nonumber\\
&  =F_{m}^{(N)}\left(  (aN)^{2}+2axN^{3/2}+x^{2}N\right)  \left[
2+2\mathrm{e}^{\frac{1}{2bN}}\cosh{\left(  \frac{x}{b\sqrt{N}}\right)
}\right] \nonumber\\
&  =F_{m}^{(N)}a^{2}N^{2}\left(  1+\frac{2x}{a\sqrt{N}}+O(N^{-1})\right)
\left[  4+O(N^{-1}\right] \nonumber
\end{align}
Comparing terms up to order $N^{-1/2}$ yields $F^{(N+1)}/F^{(N)}=(2aN)^{2}$
and $b=a^{2}/2$. With the definition of $c=4/\pi$ this yields $a=c/2=2/\pi
\approx0.637$ and $b=2/\pi^{2}\approx0.203$. This agrees to the values we
found for $N=100$ from numerically exact computation using Mathematica, apart
from finite-size corrections. So we have proved the scaling exponent $y=2$ and
also found the asymptotic form of $F_{m}^{(N)}$ except for the amplitude $F$.

\bigskip For arbitrary twisting angle $\theta$ we assume, analogously to
(\ref{AnsatzForF}),
\begin{equation}
\frac{F_{m}^{(N)}(\cot\frac{\theta}{2})^{2n}}{\sum_{m=0}^{N}F_{m}^{(N)}%
(\cot\frac{\theta}{2})^{2n}}=\frac{1}{\sqrt{4\pi b(\theta)N}}\mathrm{\exp
}\left(  -\frac{(m-a(\theta)N)^{2}}{2b(\theta)N}\right)  .
\label{AnsatzForF(Theta)}%
\end{equation}
Proceeding analogously to the lines indicated for $\theta=\pi/2$, we find
\begin{align}
\lim_{p\rightarrow0}\frac{Z(N+1,\theta)}{Z(N,\theta)}  &  =\frac{4}{\sin
^{2}\theta}a(\theta)^{2}N^{2}\label{ZratioTheta}\\
\frac{1}{2b(\theta)}  &  =\frac{2\sin^{2}\frac{\theta}{2}}{a(\theta)^{2}}.
\label{B(Theta)}%
\end{align}

\textbf{Current} $J^{z}$. We start from the expression (\ref{Jz}). Let us
denote $Q^{(N)}=\left\langle 0,0\right\vert B_{0}^{N}(-S_{z}-T_{z}))\left\vert
R_{\theta},R_{\theta}^{\ast}\right\rangle $, so that
\begin{equation}
j^{z}(N,\theta)=\frac{4}{\sin\theta}\frac{Q^{(N-1)}}{Z(N,\theta)},
\label{JzAnotherExpression}%
\end{equation}
Only diagonal matrix elements $\left\langle 0,0\right\vert ...\left\vert
n,n\right\rangle $ , $n=0,1,..N$ contribute to the $Q^{(N)}$. We are
interested in the lowest order in $p$, for which the "diagonal" part of the
vector $\left\vert R_{\theta},R_{\theta}^{\ast}\right\rangle $ becomes
\[
\left\vert R_{\theta},R_{\theta}^{\ast}\right\rangle =\left\vert
0,0\right\rangle -4p^{2}\sum_{n=1}^{\infty}\frac{\left(  \cot\frac{\theta}%
{2}\right)  ^{2n}}{n^{2}}\left\vert n,n\right\rangle +O(p^{3})+\text{nondiag.
terms}%
\]
Using\ the above, $(-S_{z}-T_{z})\left\vert n,n\right\rangle =2n\left\vert
n,n\right\rangle $, (\ref{F(N) definition}), and the recursion
(\ref{RecurrenceRelations}), we obtain for small $p$%
\begin{align*}
Q^{(N-1)}  &  =-8p^{2}\sum_{n=1}^{N-1}\frac{\left(  \cot\frac{\theta}%
{2}\right)  ^{2n}}{n}F_{n}^{(N-1)}=\\
&  =-8p^{2}\sum_{n=1}^{N-1}\left(  \cot\frac{\theta}{2}\right)  ^{2n}n\left(
2F_{n}^{(N-2)}+F_{n-1}^{(N-2)}+F_{n+1}^{(N-2)}\right)  =\\
&  =\frac{-32p^{2}}{\sin^{2}\theta}\left(  \sum_{n=1}^{N-2}n\left(  \cot
\frac{\theta}{2}\right)  ^{2n}F_{n}^{(N-2)}+\frac{\sin^{2}\theta}{4}%
F_{1}^{(N-2)}\left(  \cot\frac{\theta}{2}\right)  ^{4}\right)
\end{align*}
The last term on the rhs is of lower order in $N$ and can be neglected.
Substituting $Q^{(N-1)}$ in (\ref{JzAnotherExpression}), and using
(\ref{ZratioTheta}), (\ref{Z(N)limitSmallP}), we obtain
\begin{align*}
j^{z}(N,\theta)  &  =\frac{8}{\sin\theta}\frac{\sum_{n=1}^{N-2}n\left(
\cot\frac{\theta}{2}\right)  ^{2n}F_{n}^{(N-2)}}{Z(N-2,\theta)}\frac
{Z(N-2,\theta)}{Z(N-1,\theta)}=\\
&  =\frac{2\sin\theta}{a(\theta)^{2}N^{2}}\langle n\rangle,
\end{align*}
where $\langle n\rangle=\sum_{n=1}^{N-2}n\left(  \cot\frac{\theta}{2}\right)
^{2n}F_{n}^{(N-2)}/Z(N-2,\theta)$. Using the Ansatz (\ref{AnsatzForF(Theta)}
we get $\langle n\rangle=a(\theta)N$, to yield
\begin{equation}
j^{z}(N,\theta)=\frac{2\sin\theta}{a(\theta)N}+O\left(  \frac{1}{N^{2}%
}\right)  . \label{Jz(Theta)}%
\end{equation}

The only unknown parameter $a(\theta)$, corresponding to the maximum of the
distribution (\ref{AnsatzForF(Theta)}), on the basis of numerical evidence is
conjectured to be
\begin{equation}
a(\theta)=\frac{\sin\theta}{\theta}, \label{a(Theta)}%
\end{equation}
up to corrections of order $O(N^{-1})$. Substituting the (\ref{a(Theta)}) into
(\ref{Jz(Theta)}) and (\ref{ZratioTheta}), we obtain (\ref{RatioOfTraces1}),
and then, all steady state density profiles and steady currents
(\ref{MagnetizProfilesContinuous2}), (\ref{Jx(N)limit}), (\ref{Jz(N)limit}).

\end{document}